\documentclass[%
 reprint,
 amsmath,amssymb,
 aps,
]{revtex4-2}

\usepackage{graphicx}
\usepackage{dcolumn}
\usepackage{bm}

\usepackage{subcaption}
\usepackage{textgreek}
\usepackage{wrapfig, setspace, booktabs} 
\usepackage{lineno}
\usepackage{amsmath}
\usepackage{multirow}

\begin{document}

\preprint{APS/123-QED}

\title{Characterisation of the Beamstrahlung radiation at FCC-ee}

\author{Manuela Boscolo}
\author{Andrea Ciarma}%
\affiliation{%
 INFN - Laboratori Nazionali di Frascati, Via Enrico Fermi, 54, 00044 Frascati RM, Italy
}%


\date{\today}

\begin{abstract}
Beamstrahlung is a dominant effect 
in the beam dynamics of the high luminosity next-generation lepton collider FCC-ee. 
We characterize the beamstrahlung radiation for the beam parameters at the four working energies, from the Z-pole to the $t\bar{t}$ threshold,
and present the effect of this radiation in the Machine-Detector-Interface region. 
We discuss the conceptual need for a photon dump due to the high power produced, which is in the order of hundreds of kilowatts.
We also discuss the detector induced backgrounds due to the incoherent $e^+e^-$ pairs produced by the interaction of the beamstrahlung photons at the IP.
\end{abstract}


\maketitle


\section{\label{sec:introd}Introduction}
Future high-energy circular colliders like FCC-ee~\cite{fcc} maximise
their luminosity by squeezing the bunches at the interaction point (IP) to nanobeam scales, and by pushing the number of particles per bunch to the limit allowed by the beam-beam interaction. Due to the highly-dense bunches at the IP, the synchrotron radiation emitted during the collision in the electromagnetic field of the opposing beam, known as beamstrahlung (BS), is very intense.
Whenever a particle traverses the opposing bunch at the IP, it will radiate and may lose sufficient energy to fall below the momentum acceptance, or contribute to the formation of the high-energy tails and increase of the beam energy spread.
Similarly to the synchrotron radiation, the beamstrahlung radiation is collinear to the beam, emitted in a cone of aperture about $1/\gamma$.

We used the generator GuineaPig++~\cite{guineapig} to simulate the beam-beam interaction including the production of beamstrahlung radiation. 
This radiation was then tracked through the detector using Key4HEP~\cite{key4hep} to evaluate the induced backgrounds.
The paper is organized as follows. 
In Section~\ref{sec:bs} we describe the Beamstrahlung parameter for the FCC-ee case,
in Section~\ref{sec:charact} we
discuss the properties of the beamstrahlung radiation,
in Section~\ref{sec:dump} we introduce the first concept study for handling this intense radiation, 
in Section~\ref{sec:bhabha} we compare the Beamstrahlung radiation with the photons emitted by the Radiative Bhabha process,
in Section~\ref{sec:ipc} we describe the induced backgrounds coming from the $e^+e^-$ pairs produced by the beamstrahlung photons, 
and finally in Section~\ref{sec:conclusions} we give few concluding remarks.

\section{\label{sec:bs} Beamstrahlung radiation}

The strength of the BS radiation is characterized by the dimensionless beamstrahlung parameter $\Upsilon$~\cite{yoyoka}, defined as

\begin{equation}
    \Upsilon = \frac{2}{3}\frac{\hbar \omega_c}{E}
\end{equation}

\noindent where $E$ is the particle energy before the radiation emission and $\hbar\omega_c=\frac{3}{2}\frac{\hbar c\gamma^3}{\rho}$ is the critical energy $E_{crit}$ \cite{Sands:1969lzn} for a particle with curvature radius $\rho$.
Each bunch particle experiences a different beam-beam force depending on its position in the 6D phase space, and it also varies during the crossing of the opposing bunch, so the BS parameter $\Upsilon$ is not constant during the collision.
The maximum and average values $\Upsilon_{ave}$
and $\Upsilon_{max}$ are used to describe the global behaviour of the two colliding bunches.
For head-on collisions of Gaussian bunches these values are defined as:

\begin{equation}
\Upsilon_{ave} \sim \frac{5}{6}\frac{r_e^2\gamma N_e}{\alpha\sigma_z(\sigma_x^{*}+\sigma_y^{*})} \ \ \ \ \ \Upsilon_{max}=\frac{12}{5}\Upsilon_{ave}
\label{eq:upsilon}
\end{equation}

\noindent where $\alpha$ is the fine structure constant, $r_e$ the electron classical radius, $\gamma$ the Lorentz factor, $N_e$ the bunch population,  $\sigma_{x,y}^*$ the r.m.s. bunch dimensions at the collision point, and $\sigma_z$ the longitudinal bunch length. 
It appears from Equation~\ref{eq:upsilon} that $\Upsilon$ is linearly proportional to the beam energy, and to the bunch density.
For FCC-ee it has been found in simulations~\cite{prab_dmitry} that the BS has a strong effect on the beam lifetime and luminosity, being the main driver for the energy spread and bunch length.
The derivation of the equilibrium emittances and bunch length in presence of  BS for lepton circular colliders is described in~\cite{Garcia:2016IPAC}, resulting in agreement with the numerical simulations performed so far~\cite{Shatilov}.
At the FCC-ee Z-pole the equilibrium bunch length and energy spread are increased by about a factor three with respect to the non-colliding bunches (i.e. dominated by the synchrotron radiation only)~\cite{oide-fccis}.

Table~\ref{tab:bs} reports $\Upsilon_{ave}$ evaluated using Equation~\ref{eq:upsilon} for various circular and linear $e^+e^-$ colliders~\cite{slc,ilc,clic,nlc,skekb}.
For circular colliders the equilibrium beam parameters have been considered.

\begin{table}[ht!]
\centering
 \caption{Relevant parameters for the BS description for various circular and linear $e^+e^-$ colliders.}
\begin{tabular}{ r  |  ccccc | l}
\hline \hline
   &E & $N_e$      & $\sigma_x^*$ & $\sigma_y^*$ & $\sigma_z$      & $\Upsilon_{ave}$\\
   & [GeV]& [$10^{10}$] & [$\mu$m]     & [nm]    & [mm]         & [$1$]\\
 \hline 
FCCee Z-pole    & 45.6 & 17 & 6.4 & 28 & 12.1 & $175\times10^{-6}$\\
FCCee $t\bar{t}$ & 182.5& 23 & 38.2 & 68 & 2.54  & $760\times10^{-6}$\\
SuperKEKB $e^-$ & 7 & 9.04 & 10.1 & 48 & 6 & $18\times10^{-6}$\\
SuperKEKB $e^+$ & 4& 6.53 & 10.7 & 62 & 5  & $86\times10^{-6}$\\
\hline
SLC         & 50& 4 & 2.1 & 900 & 1.1  & 0.00152 \\
ILC         & 250& 2 & 0.64 & 5.7 & 0.3  & 0.0456\\
CLIC        & 165& 0.52 & 0.149 & 2.9 & 0.07  & 0.144\\
CLIC 3TeV   & 1500& 0.37 & 0.045 & 1.0 & 0.044  & 4.91\\
NLC         & 1000& 1.1 & 0.36 & 2.3 & 0.1  & 0.535\\
 \hline \hline
\end{tabular}
\label{tab:bs}
\end{table}

At FCC-ee $\Upsilon_{ave} \ll 1$ and beams are flat, as $\sigma_y / \sigma_x\ll 1$.
Under these conditions the particle average energy loss $\delta_{E}$ can be expressed as a function of $\Upsilon_{ave}$:

\begin{equation}
\label{eq:deltae}
    \delta_E = \frac{16\sqrt{3}}{5\pi^{3/2}}\frac{r_e\alpha N_e}{\sigma_x^{*}}\Upsilon_{ave} = \frac{24}{3\sqrt{3}\pi^{3/2}}\frac{r_e^3\gamma N_e^2}{\sigma_z\sigma_x^{*2}}
\end{equation}

The average number of photons emitted per particle, $n_\gamma$, and their average energy, $<E_\gamma>$, can be evaluated according to:

\begin{equation}
\label{eq:ngamma}
n_\gamma = \frac{12}{\pi^{3/2}}\frac{\alpha^2\sigma_z}{r_e\gamma}\frac{6}{5}\Upsilon_{ave} = \frac{12}{\pi^{3/2}}\frac{\alpha r_e N_e}{\sigma_x^{*}}
\end{equation}

\begin{equation}
\label{eq:egamma}
<E_\gamma> = \frac{\delta_E}{n_\gamma}E_e=\frac{4\sqrt{3}}{15}\Upsilon_{ave}E_e
\end{equation}

Equations~\ref{eq:deltae}, \ref{eq:ngamma} and \ref{eq:egamma} indicate the dependence 
 of the bunch density on the BS effect.

FCC-ee adopts the crab-waist collision scheme~\cite{crabw} that requires a large 
Piwinski angle, defined by
$\Phi = \sigma_z / \sigma_x^{*} tan\left(\theta/2\right)$, where $\theta$ is the horizontal crossing angle. This scheme modifies the effective length of the interaction area, which can be defined as:

\begin{equation}
    L_i = \frac{\sigma_z}{\sqrt{1+\Phi^2}}
\end{equation}

For Piwinski angles above $\Phi\geq1$, $L_i$ is smaller and not comparable with the bunch length.
In this case Eqs.~\ref{eq:deltae} and \ref{eq:ngamma} become~\cite{Garcia:2019nci}:

\begin{equation}
\label{eq:ngamma2}
n_\gamma = \frac{12}{\pi^{3/2}}\frac{\alpha r_e N_e}{\sigma_x^{*}} \frac{1}{\sqrt{1+\Phi^2}}
\end{equation}

\begin{equation}
\label{eq:deltae2}
    \delta_E = \frac{24}{3\sqrt{3}\pi^{3/2}}\frac{r_e^3\gamma N_e^2}{\sigma_z\sigma_x^{*2}} \frac{1}{\sqrt{1+\Phi^2}}
\end{equation}

\section{\label{sec:charact} Beamstrahlung radiation power and divergence}

We characterised the BS radiation at FCC-ee with GuineaPig++ simulations, referring to the beam parameters in Table~\ref{tab:4iplattice}.
The beams are assumed Gaussian because we are interested in estimating the total power emitted via this process, so photons emitted by particles outside the core have a negligible contribution.

\begin{table*}[hbt!]
\centering
\caption{Beam parameters for the FCC-ee 4 IP lattice v530 as from Ref.~\cite{oide-fccis}. 
The index SR refers to the single beam equilibrium parameters in presence of synchrotron radiation only, the index BS refers to the collision case, with beamstrahlung.
The last two rows indicate the total power and mean energy of the beamstrahlung radiation, as obtained from GuineaPig simulations.}
\label{tab:4iplattice}
\begin{tabular}{ l  l  |  c c c c }
\hline \hline
   Parameter & Units & Z & WW & ZH & $t\bar{t}$\\ 
 \hline 
beam energy & GeV & 45.6 & 80.0 & 120.0 & 182.5 \\
horizontal emittance $\epsilon_x$ & nm rad & 0.71 & 2.16 & 0.64 & 1.49 \\
vertical emittance  $\epsilon_y$ & pm rad & 1.42 & 4.32 & 1.29 & 2.98 \\
horizontal IP  $\beta_x$ & m & 0.1 & 0.2 & 0.3 &1 \\
vertical IP $\beta_y$ & mm & 0.8 & 1 & 1 & 1.6 \\
horizontal IP beam size $\sigma^*_x$ & $\mu$m & 8.46 & 20.78 & 13.86 & 38.60  \\
vertical IP beam size $\sigma^*_y$ & nm & 33.7 & 65.7 & 35.9 & 69.0 \\
bunch length $\sigma_z$ (SR/BS) & mm & 4.38/15.4 & 3.55/8.01 & 3.34/6.0 & 2.00/2.74 \\
momentum spread $\sigma_\delta$ (SR/BS) & \% & 0.038/0.132 & 0.069/0.154 & 0.103/0.185 & 0.157/0.221 \\
bunch population $N_e$ & $10^{11}$ & 2.43 & 2.91 & 2.04 & 2.37 \\
bunches/beam $n_{b}$ &  & 10000 & 880 & 248 & 40 \\
\hline
BS total power & kW & 370 & 250 & 150 & 80 \\
BS mean energy $<E_\gamma>$ & MeV & 2 & 7 & 23 & 63 \\
 \hline \hline
\end{tabular}
\end{table*}

In the last two rows of Table~\ref{tab:4iplattice} we list the resulting total power and the mean energy of the produced BS photons, obtained from GuineaPig++ simulations. 
The highest power value is about 400\,kW and is observed at the Z-pole due to the high current at this energy. 
The average radiation energy is highest at the $t\bar{t}$, as expected also from Eq.~\ref{eq:deltae}.
Photon mean energies range from 2 to 63\,MeV, with tails of few hundreds of MeV at the Z-pole and up to few GeV at the $t\bar{t}$ threshold. 
The peak photon flux goes from $10^{12}$ to $10^{14}$ photons/s per 0.1\% bandwidth, growing with the working point energy, due to the beam current.
These results are in very good agreement with the first estimates discussed in Refs.~\cite{hbu_bs,Boscolo:2021hsq}.

\begin{figure}[!hbt]
    \centering
    	\includegraphics*[width=0.48\textwidth]{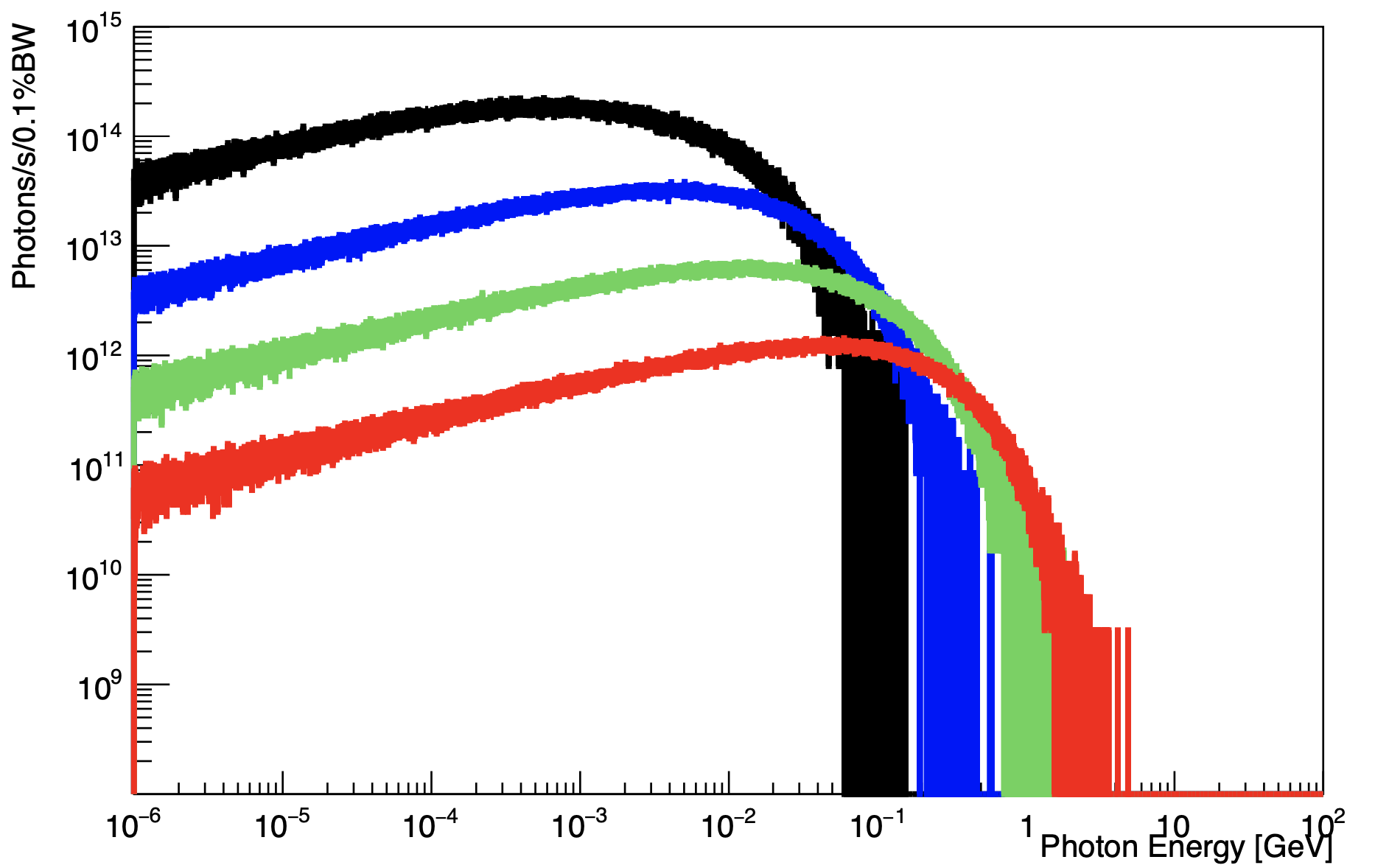}
    \caption{Flux of the BS radiation as a function of their energy, emitted for the four FCC-ee working points, 45.6\,GeV (black), 80.0\,GeV (blue), 120.0\,GeV (green), and 182.5\,GeV (red).
    }
    \label{fig:spectra1}
\end{figure}

The flux of the BS photons per unit of bandwidth as a function of their energy at the four working points is shown in Figure~\ref{fig:spectra1}.

This intense and energetic photon flux requires to be properly handled.
We discuss in the following the first estimates to evaluate the photon spot size and divergence, necessary to establish possible locations of the BS dump in the tunnel. 
These evaluations have been used to consider the feasibility of an extraction photon beam line for a dump at few hundreds meters from the IP.

Table~\ref{tab:angular} reports the photon and electron beam divergences, which result comparable, as expected. 
The photon angular spread goes from about 45\,$\mu$rad at the $t\bar{t}$ threshold to about 92\,$\mu$rad at the Z-pole. 
These values correspond to a
photon beam spot of about 1\,$\rm cm^{2}$ at 100\,m downstream the collision point at the Z-pole, and about 0.5\,$\rm cm^{2}$ at the $t\bar{t}$ threshold.

\begin{table}[h!bt]
\centering
\begin{tabular}{ r  |  cccc}
\hline \hline
 & $\sigma_{px}(\gamma)$ & $\sigma_{py}(\gamma)$ & $\sigma_{px}(e^-)$ & $\sigma_{py}(e^-)$ \\
 & [$\mu$rad] & [$\mu$rad] & [$\mu$rad] & [$\mu$rad] \\
\hline 
Z & 91.8 & 49.2 & 84.3 & 42.1 \\
WW & 110 & 73.0 & 103.4 & 65.7 \\
ZH & 51.7 & 41.3 & 46.2 & 35.9 \\
$t\bar{t}$ & 44.6 & 50.3 & 38.6 & 43.2 \\
 \hline \hline
\end{tabular}
 \caption{Divergence of the photon and electron beams.} 
 \label{tab:angular}
\end{table}

\begin{figure}[b]
    \centering
    \includegraphics[width=246pt]{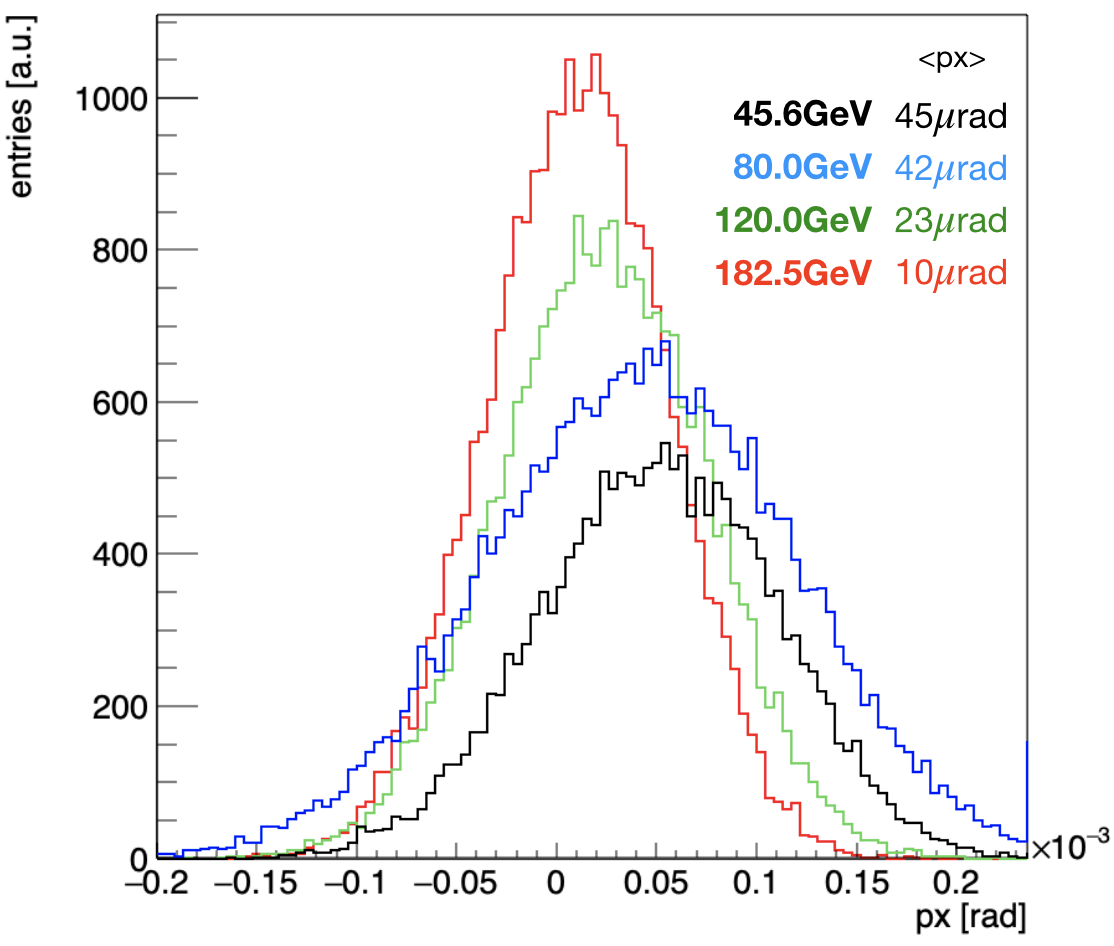}
    \caption{Horizontal angle distribution of the BS radiation for the four FCC-ee working points.
    }
    \label{fig:rigidity}
\end{figure}

Figure~\ref{fig:rigidity} shows the horizontal angular distribution of the BS radiation at the four FCC-ee working points.
Photons emitted at higher energies have the center of the distribution closer to the longitudinal beam axis (corresponding to zero in the plot).
This can be explained by the consideration that higher energy particles are more rigid and curve less in a magnetic field, according to the definition of beam rigidity $B\rho = P/0.3$, where $P$ is the particle momentum expressed in GeV/c.
The difference between the values of $<p_x>$ at the lowest and highest energies (Z-pole and $t\bar{t}$ threshold) is 35\,$\mu$rad.
Comparing these values with the photon angular divergences in Table~\ref{tab:angular}, it can be observed that this offset is small enough to assure that the cores of the distributions are still broadly overlapped, as appears in Figure~\ref{fig:rigidity}.

\begin{figure}[!hbt]
    \centering
    \includegraphics[width=246pt]{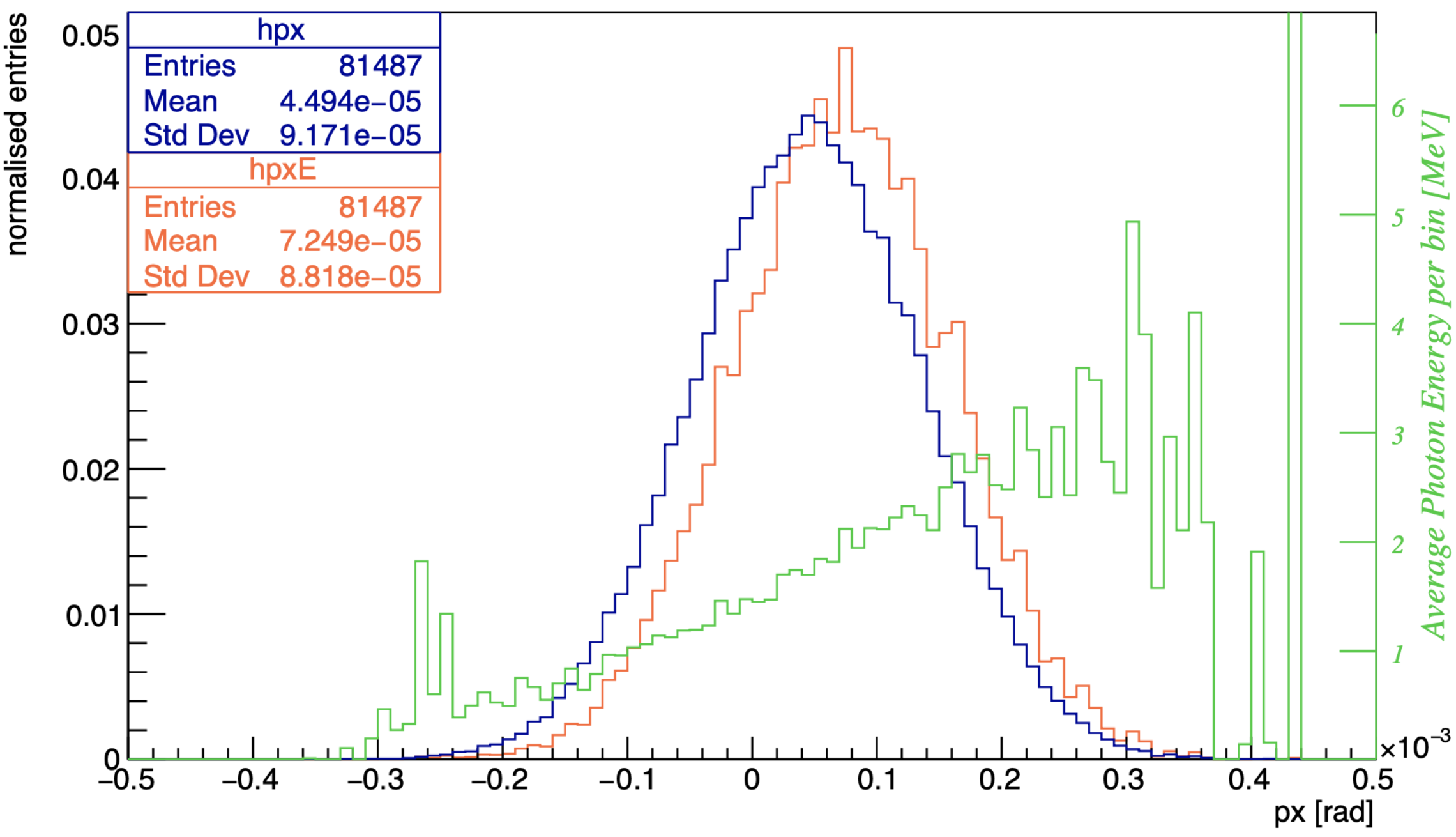}
    \caption{Comparison of the geometrical (blue) and weighted for the energy (red) horizontal angular distribution for the BS photons emitted at the Z-pole. In green the average photon energy per each angular bin.}
    \label{fig:dmitry}
\end{figure}

Figure~\ref{fig:dmitry} shows the correlation between the BS photon energy (in blue) and the horizontal emission angle (in red). 
The energy of the BS photon depends on the intensity of the beam-beam kick received by the emitting particle.
This appears by the green histogram of Figure~\ref{fig:dmitry}, which reports the average energy of the photons emitted at a given horizontal angle.
If the angular distribution of the BS radiation is weighted by 
the photon energy, we notice that 
the center of the power angular distribution is slightly shifted with respect to the geometrical one.
The difference between the power and the geometrical angular distribution 
is about 10\,$\mu$rad, implying that even at few hundreds meters from the IP, the offset would be about 1\,mm for the four working points.

Therefore, even applying an offset due to different magnetic rigidity and photon energy distribution, the spots will be mostly overlapping at all beam energies.

\begin{figure}[!hbt]
    \centering
    \includegraphics[width=246pt]{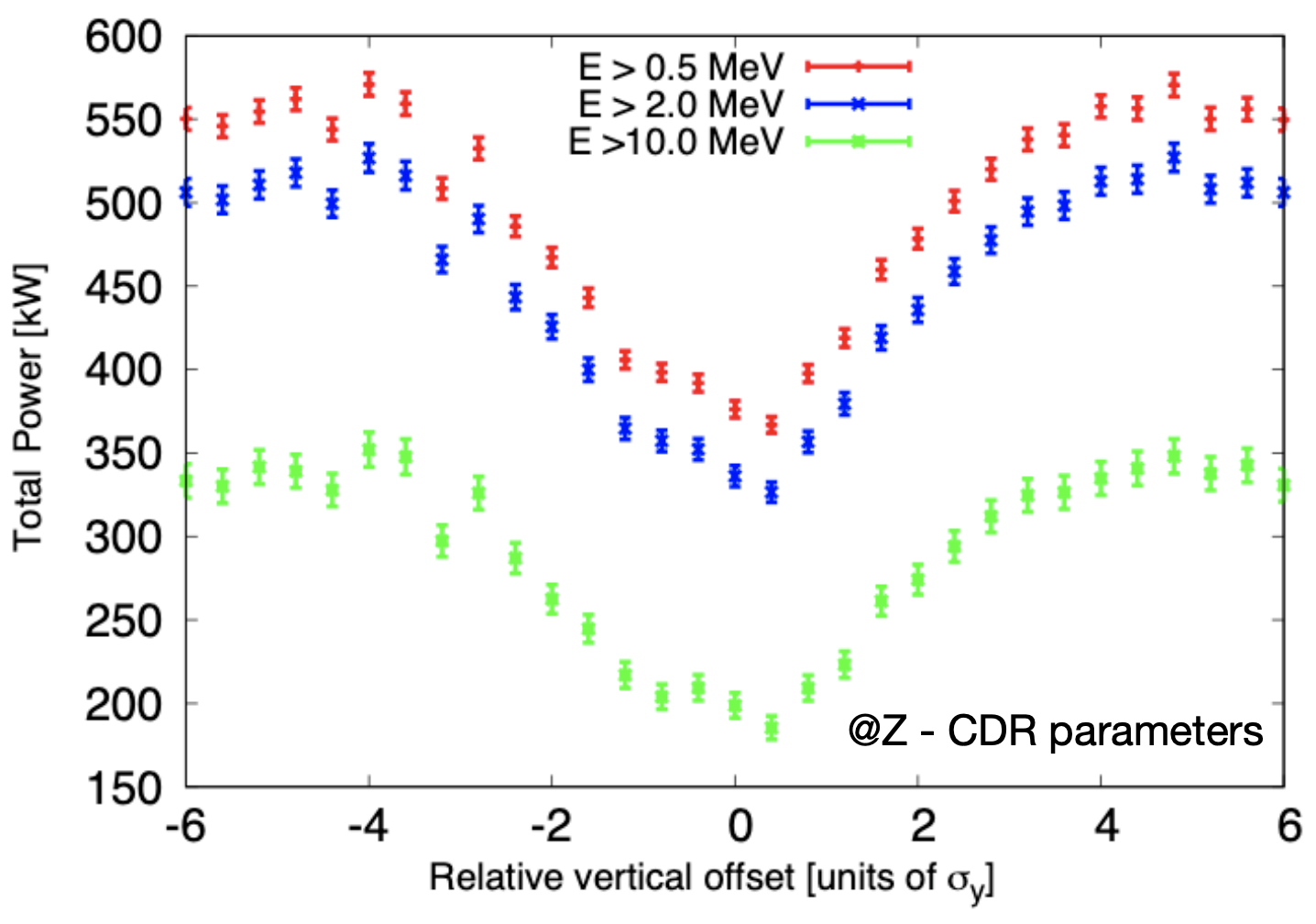}
    \caption{Total power of the BS radiation as a function of the vertical offset between the colliding beams with beam parameters at the Z-pole. Different energy cuts on the photon energy are shown in red, blue and green.}
    \label{fig:yoffset}
\end{figure}

The strength of the beam-beam force varies with the offset between the colliding bunches, therefore also the intensity of the BS radiation is strongly dependent on this parameter.
We performed a scan on the vertical offset between the bunches.
Figure~\ref{fig:yoffset} shows the result of this study at the Z-pole. 

In the event of two bunches colliding with an offset, the total radiated power at the Z-pole increases to values that exceed 500\,kW for lattice v530. 
We applied three energy cuts, 0.5, 2.0, and 10\,MeV to analyse the photon energy as a function of the vertical offset, shown in red, blue and green curves respectively.
It can be observed that about half of the power is carried by hard photons,
in agreement with the definition of the critical energy as for the synchrotron radiation, which is related to the mean photon energy by $E_{crit} = 15\sqrt{3}/8<E>$. $E_{crit}$ is about 7\,MeV at the Z-pole.

\section{\label{sec:dump} Beamstrahlung photon beam trajectory}
\begin{figure*}[!hbt]
    \centering
    \includegraphics[width=0.9\textwidth]{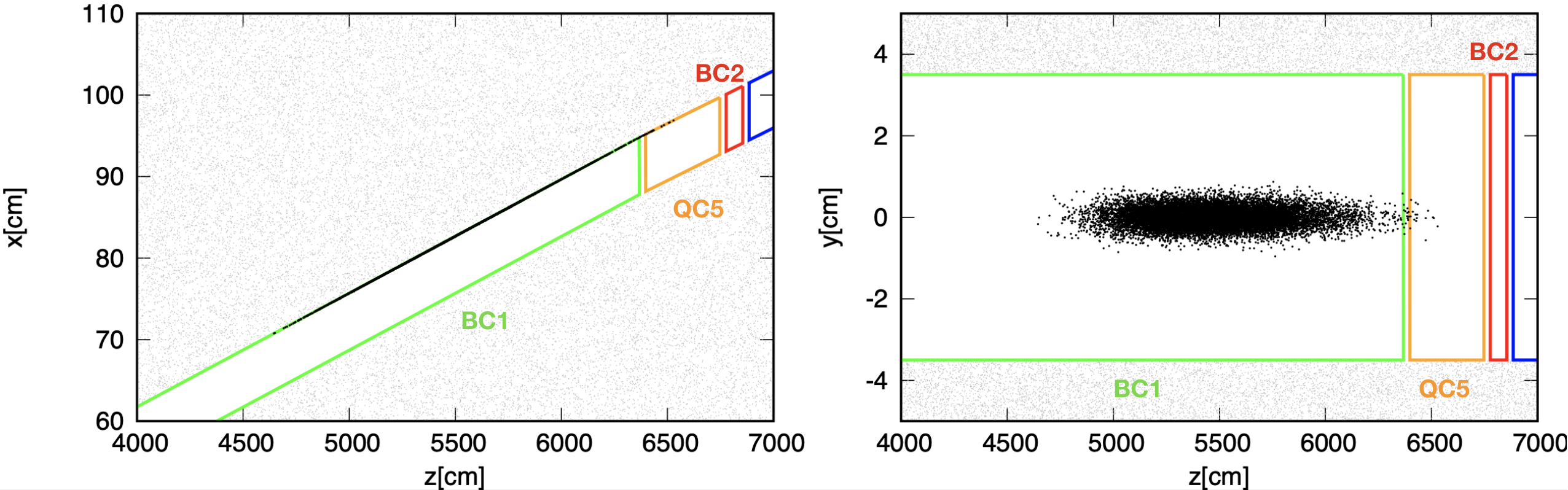}
    \caption{Left: Photons (black dots) hitting the vacuum chamber inside the first bending magnet (BC1) after the IP for the Z-pole c.o.m. energy. Right: corresponding vertical spot as a function of the longitudinal coordinate. The longitudinal distance is referred to the IP.}
    \label{fig:tracking}
\end{figure*}

The BS radiation is collinear with the beam, as discussed in section~\ref{sec:charact}, so  it will hit the vacuum chamber of the first downstream bending magnet (BC1), which is  40\,m long and starts about 25\,m from the IP.
Figure~\ref{fig:tracking} shows the photons, depicted in black dots, hitting the internal side of the vacuum chamber of BC1 for the Z-pole c.o.m. energy case.
Although the transverse spot size of the photons is about 1\,$\rm cm^2$,
the irradiated region is several meters long due to the angle of incidence of the photons, which is about 1\,mrad.

\begin{figure}[!hbt]
    	\centering
    	\includegraphics[width=0.48\textwidth]{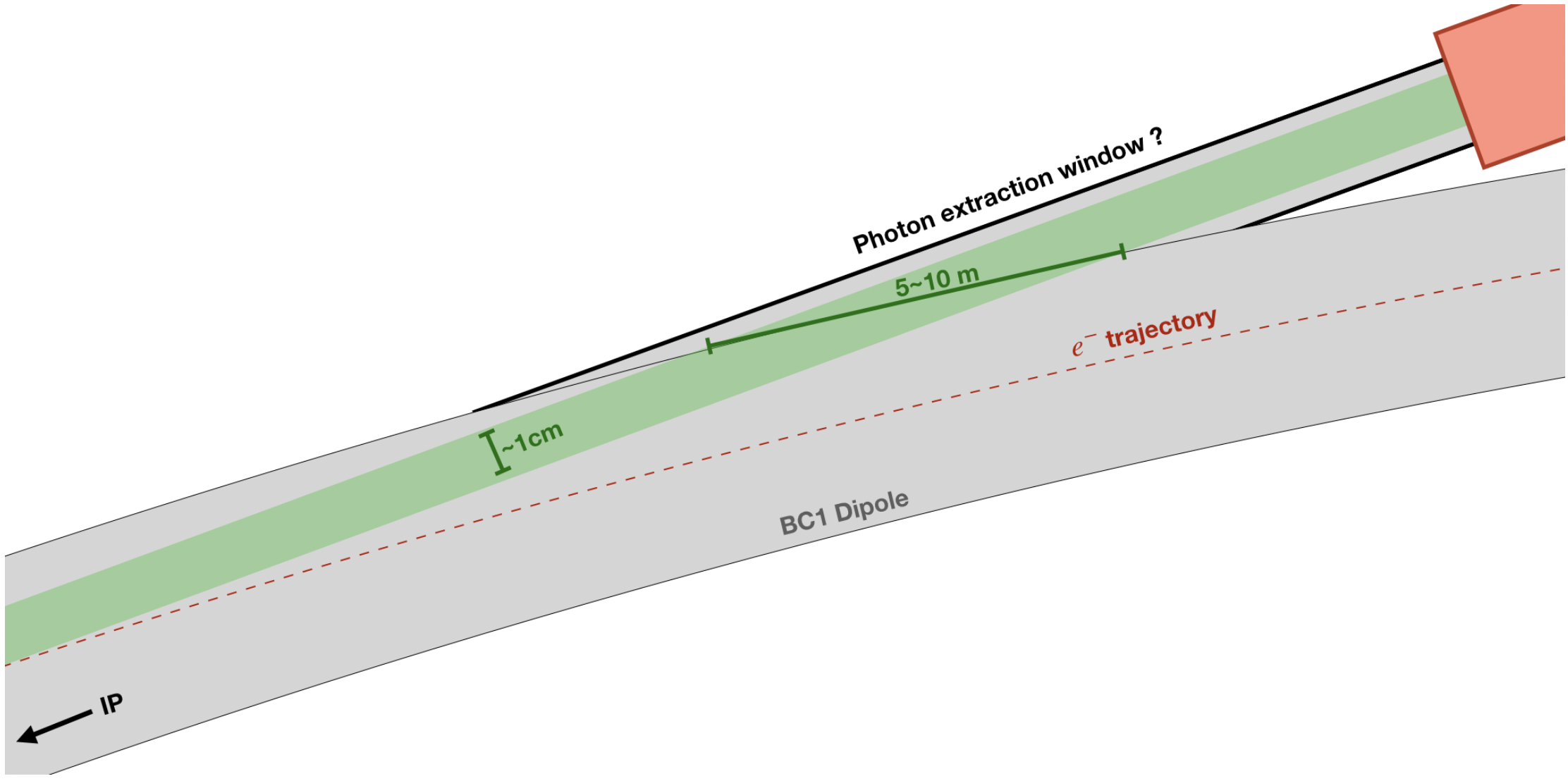}
 \caption{Conceptual sketch of a BS photon extraction line at the first bending magnet after the IP (not to scale).}
    \label{fig:extraction1}
\end{figure}

Figure~\ref{fig:extraction1} shows the conceptual design of the BS extraction line.
First evaluations performed by the magnet experts are encouraging and confirm the possibility to design custom magnets downstream the IP
with large aperture yokes to extract the BS radiation from the beam vacuum chamber and transport it to a proper dump~\cite{jeremie}.

\begin{figure}[!hbt]
    	\centering
    	\includegraphics[width=0.45\textwidth]{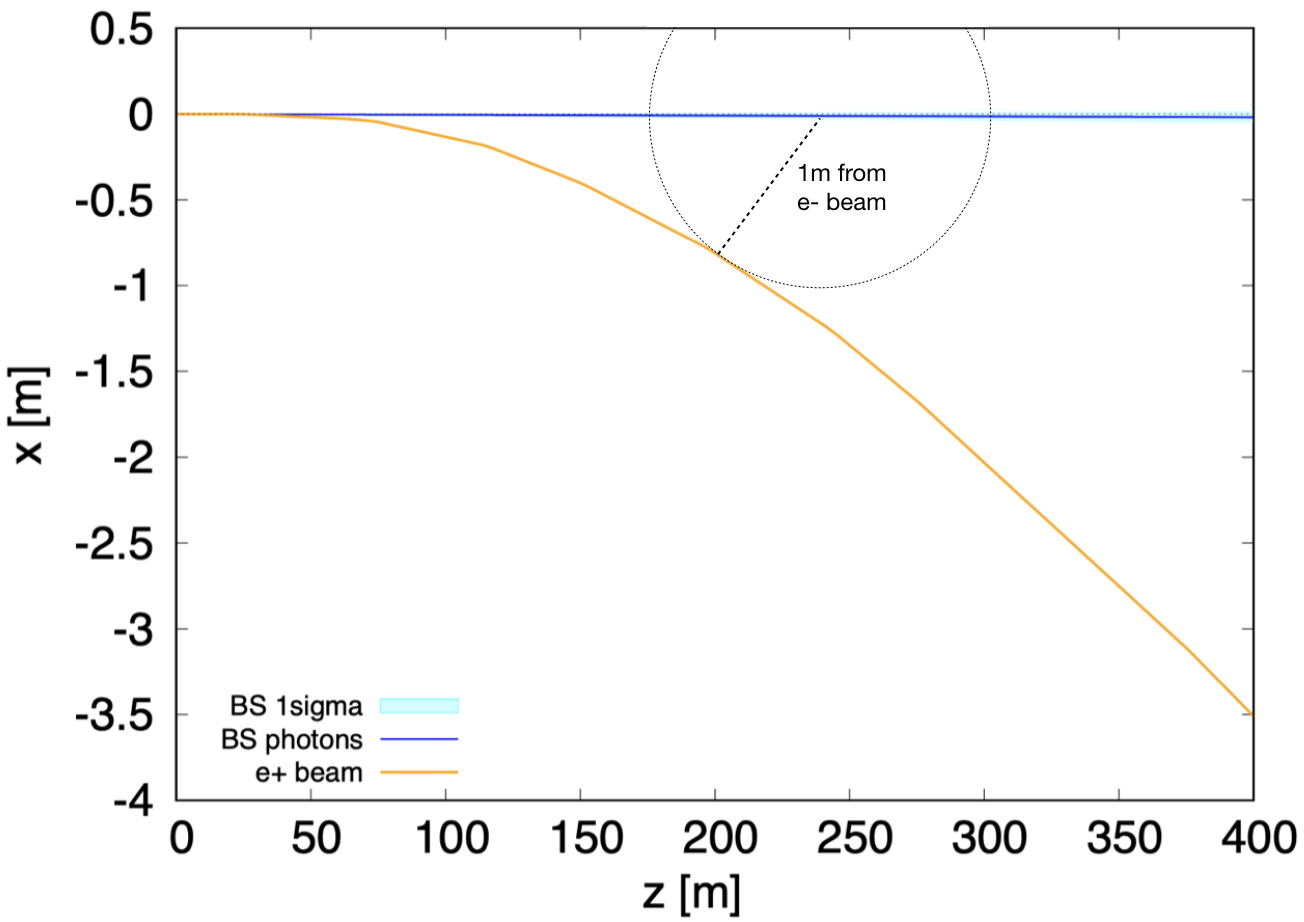}
\caption{Electron/positron and BS photon beam lines. At 250\,m after the IP the distance is 1\,m}
    \label{fig:extraction2}
\end{figure}

Comparing the trajectory of the positron (or electron) beams obtained from MAD-X~\cite{madx} with that of the BS radiation, we observe that at about 250\,m from the IP the two beam lines are at an horizontal distance of 1\,m, as shown in Figure~\ref{fig:extraction2}.
This distance is about the minimum required to position a separate alcove for a BS photon dump.
According to these studies, we agreed with the civil engineering group to set the BS dump at about 500\,m from the IP~\cite{fani}.

We note that in a 4 IP lattice it will be necessary to have a total of 8 photon extraction lines and beam dumps, two for each IP. 

\subsection{Dependence of BS radiation to beam parameters and optics}

In this paragraph we compare the BS radiation described up to now for optics v530 with the one resulting from beam parameters of optics v566~\cite{v566} (see Table~\ref{tab:newtable}).

The number of photons emitted per particle and their average energy are proportional to the beamstrahlung parameter, which in turns is proportional to the bunch population according to Equations~\ref{eq:upsilon}, \ref{eq:ngamma} and \ref{eq:egamma}. 
The total BS power for v566 is about a factor 2 lower than for v530, but still in the order of 100\,kW.

\begin{table}[h!bt]
\centering
\begin{tabular}{ rr  |  cccccc}
\hline 
  & & $N_e$ & Tot. Power & $<E_\gamma>$ & $<n_\gamma>$ & $<p_x> $ & $\sigma_{p_x}$ \\
  & & [$10^{11}$] & [kW] & [GeV] & [1] & [$\mu$rad] & [$\mu$rad] \\
\hline \hline
\multirow{4}{*}{\rotatebox[origin=c]{90}{v530}} 
 & Z & 2.43 & 370 & 2 & 0.16 & 45 & 91 \\
 & WW & 2.91 & 236 & 7 & 0.25 & 35 & 110 \\
 & ZH & 2.04 & 147 & 22 & 0.25 & 23 & 52 \\
 & $t\bar{t}$ & 2.37 & 77 & 62 & 0.24 & 10 & 45 \\
 \hline \hline
 \multirow{4}{*}{\rotatebox[origin=c]{90}{v566}}
 & Z & 1.51 & 181 & 1.5 & 0.11 & 30 & 84 \\
 & WW & 1.47 & 118 & 6 & 0.16 & 20 & 107 \\
 & ZH & 1.15 & 80 & 19 & 0.16 & 16 & 57 \\
 & $t\bar{t}$ & 1.55 & 41 & 53 & 0.16 & 5 & 43 \\
 \hline 
\end{tabular}
 \caption{BS parameters for v530~\cite{oide-fccis} and v566~\cite{v566} optics.} 
 \label{tab:newtable}
\end{table}

The reduction of the number of particles per bunch not only reduces the BS radiation power, but also the intensity of the beam-beam kick, as well as the mean value of the photon horizontal momentum. 
This difference between the two optics is small enough to assure that the core of the two beam distributions are broadly overlapped.

On the other hand, the RMS of the angular distributions does not change significantly, as this value is dominated by the divergence of the electron (positron) beam.

\begin{figure*}[!hbt]
    	\centering
    	\includegraphics[width=0.7\textwidth]{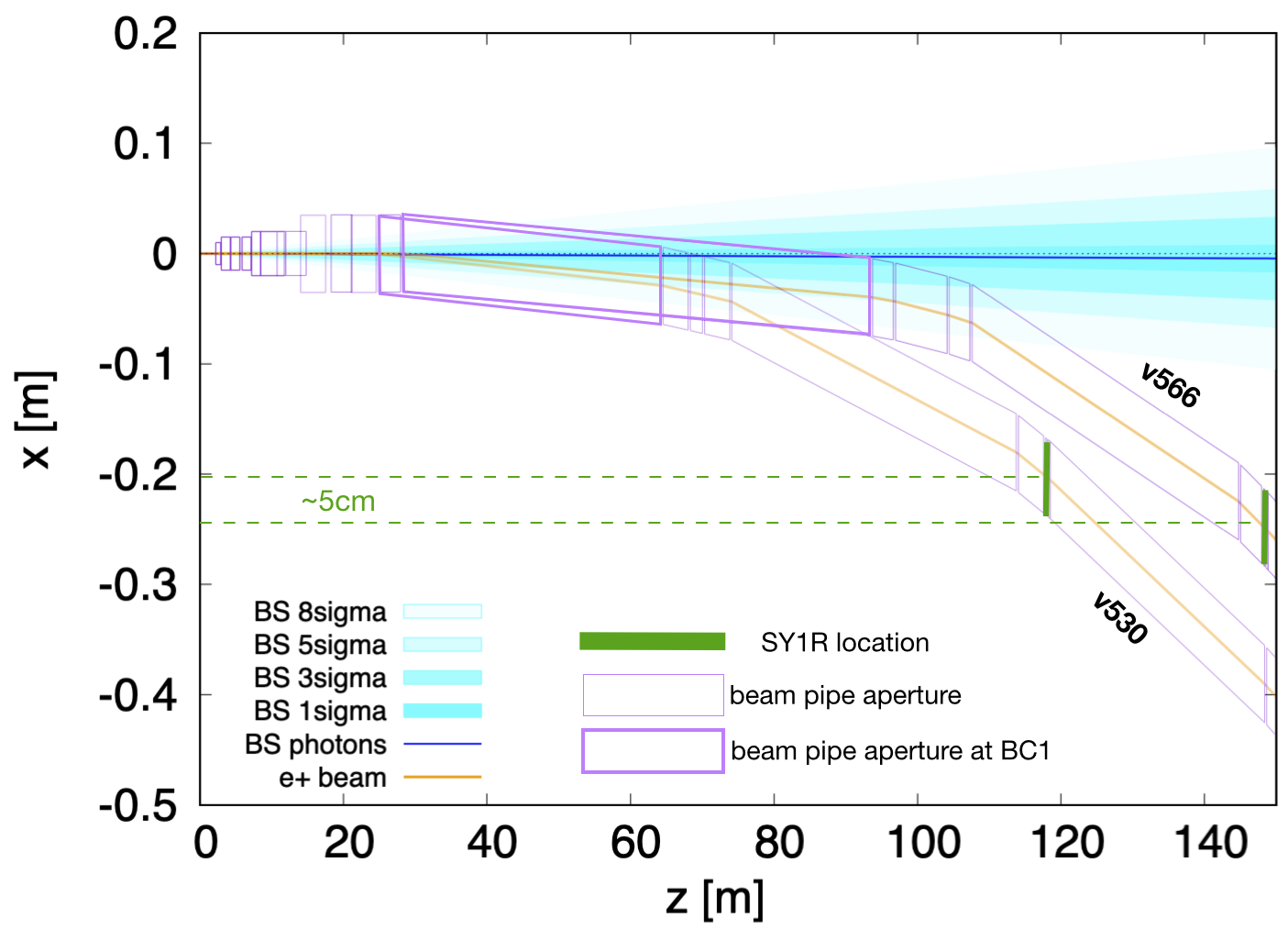}
\caption{Comparison of the beam lines after the IP for v530 and v566 optics.}
    \label{fig:survey_comparison}
\end{figure*}

Figure~\ref{fig:survey_comparison} shows the comparison of the two beam lines downstream the IP for the v530 and v566 optics. 

For the v566 case the first dipole after the IP (BC1) is much longer than for v530.
This modification is relevant afor the design of the dipoles between the IP and the BS extraction line. 
The aperture yokes of the dipoles need to be larger than the BS photon envelope, and  
the BS photon envelope is strongly dependent on the dipoles location, magnetic length and strength.

\section{\label{sec:bhabha} Photons from radiative Bhabha scattering}

An additional source of radiation coming from the IP is radiative Bhabha (otherwise known as Single Beamstrahlung).
This radiation is emitted in the same radiation cone of about $1/\gamma$ as for the BS,
so this radiation will be extracted and dumped in the same BS dump.

We used the Monte Carlo generator BBBREM~\cite{BBBREM} and GuineaPig++ to characterize the radiation emitted via this process at the four FCC-ee working points.

\begin{figure*}[!hbt]
    	\centering
    	\includegraphics[width=0.8\textwidth]{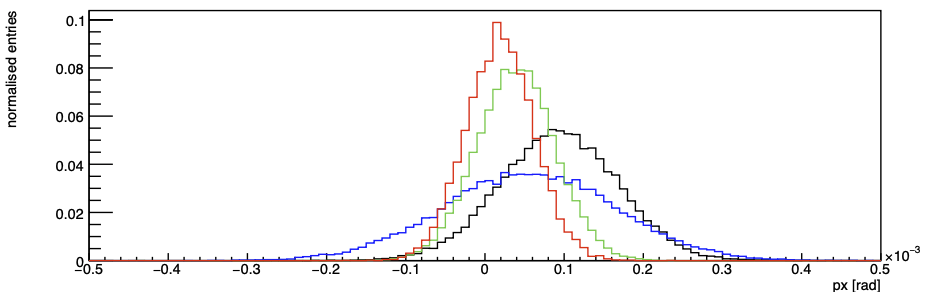}
 \caption{Radiative Bhabha horizontal momentum distribution at the four FCC-ee working points: Z (black), WW (blue), ZH (green), $t\bar{t}$ (red).}
    \label{fig:RBpx}
\end{figure*}

Figure~\ref{fig:RBpx} shows the horizontal momentum distribution in the downstream beam pipe reference frame for the radiative Bhabha photons at the four FCC-ee working points.
The photon mean horizontal momentum is inversely proportional to the magnetic rigidity of the beam (i.e. the nominal energy).
These values are comparable with those obtained for the BS radiation (Figure~\ref{fig:rigidity}), so that the two photon beams will overlap.
Table~\ref{tab:angularRB} shows the total power carried by the radiative bhabha photons, together with the average and RMS values of the radiation horizontal momentum.

\begin{table}[h!bt]
\centering
\begin{tabular}{ r  |  ccc}
\hline \hline
 & $P_{tot}$ &$<p_x>$ & $\sigma_{p_x}$  \\
 & [W] & [$\mu$rad] & [$\mu$rad]  \\
\hline 
Z  & 430 &	93	& 76 \\
WW & 60 &	56	& 109 \\
ZH & 40 &	39	& 50 \\
$t\bar{t}$ & 2 & 19 & 43 \\
 \hline \hline
\end{tabular}
 \caption{Divergence of the radiative Bhabha photon beam at the four working points.} 
 \label{tab:angularRB}
\end{table}

\begin{figure}[!hbt]
    	\centering
    	\includegraphics[width=0.48\textwidth]{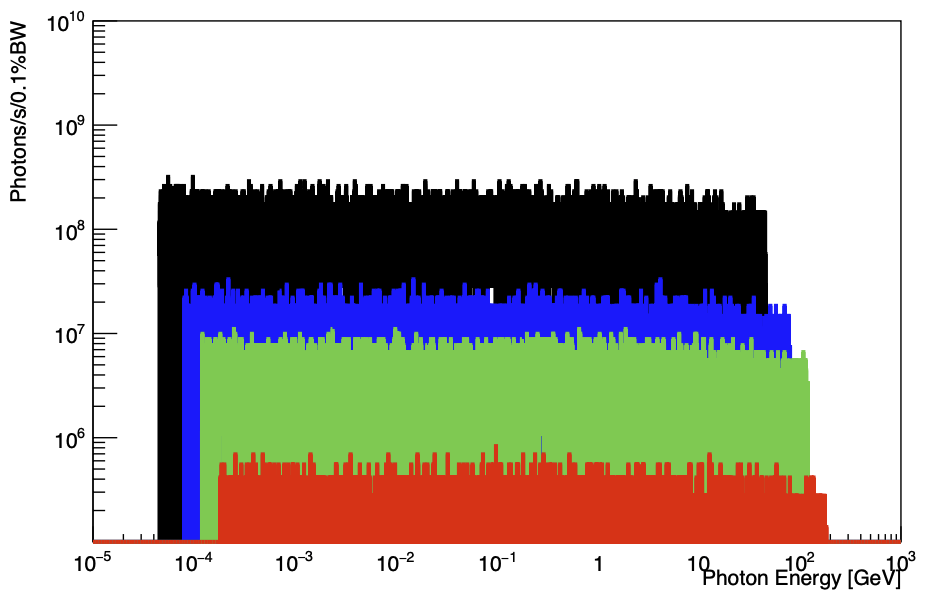}
 \caption{Radiative Bhabha energy spectrum at the four FCC-ee working points: Z (black), WW (blue), ZH (green), $t\bar{t}$ (red).}
    \label{fig:RBspectrum}
\end{figure}

Figure~\ref{fig:RBspectrum} shows the energy spectrum of the radiative Bhabha photons at the four FCC-ee working points.
Compared to the beamstrahlung energy spectrum (Figure~\ref{fig:spectra1}) this radiation is several order of magnitudes less intense, but the endpoint energy of the distribution corresponds to the nominal energy of the beam, while for the beamstrahlung it is few MeV.

In terms of radiated power, the highest value is obtained at the Z-pole for $\mathcal{O}(500W)$, going down to few Watts at the other working points.
Therefore the contribution to the deposited power on the beamstrahlung dump due to radiative Bhabha can be considered negligible.

This process has been used in several lepton colliders to provide fast luminosity measurements, like DA$\Phi$NE, VEPP, PEP-II, \cite{rb_dafne,rb_vepp,rb_pep2}.
The possibility to use radiative Bhabhas for relative luminosity measurements for machine tuning is 
being considered. The challenge here is to be able to discriminate radiative Bhabhas from BS photons as well as SR.

\section{\label{sec:ipc} Beamstrahlung induced detector backgrounds}

The beamstrahlung radiation is produced in the direction of the outgoing beams and does not directly hit the detector.
However, the interaction of the beamstrahlung photons with other real or virtual photons at the IP can be the source of $e^+e^-$ pairs emitted at large angles, which can enter the detector and cause background.
This process is known as Incoherent Pair Creation (IPC)~\cite{yoyoka}.

\begin{figure*}[!hbt]
    \centering
    \includegraphics[width=0.8\textwidth]{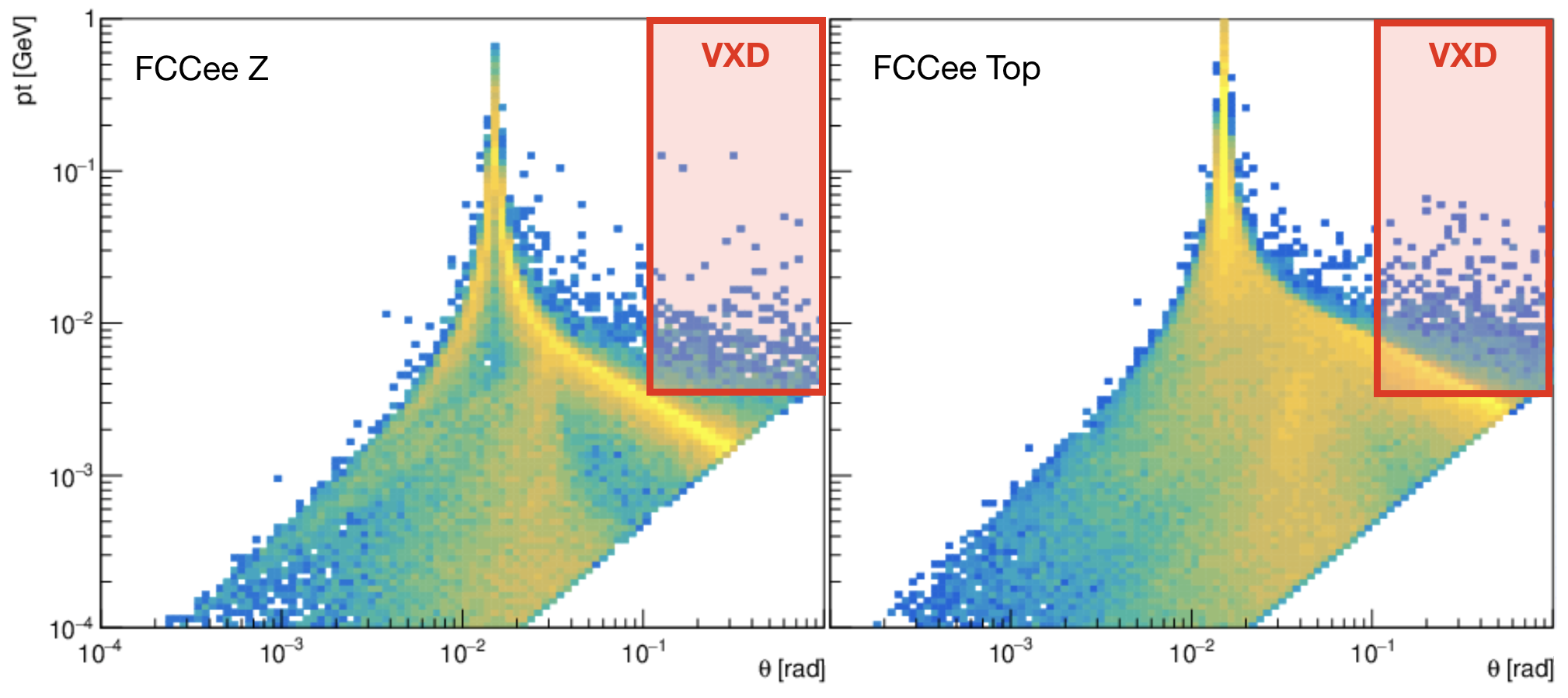}
    \caption{Production kinematics of the IPC particles emitted at FCC-ee for the Z and Top working points. The red area represents the acceptance of the CLD vertex detector (VXD).}
    \label{fig:ipc_kin}
\end{figure*}

Figure~\ref{fig:ipc_kin} shows the production kinematics of these particles for the Z and $t\bar{t}$ working points at FCC-ee.
Most of the high-energy particles are produced at an angle of $\theta=0.015$\,rad in the detector reference frame.
This value means that these particles are emitted collinearly with the beam.

For $p_T<$100\,MeV two distinct particle families can be identified according to their deflection angle.
Particles emitted in the direction of the beam with the opposite charge (e.g. a positron from the pair emitted in the direction of the outgoing electron beam) will experience a focusing force due to the electric field and will populate the two branches at low $\theta$ visible in Figure~\ref{fig:ipc_kin}.
On the other hand, particles emitted in the direction of the beam with the same charge (e.g. a positron from the pair emitted in the direction of the outgoing positron beam) will experience a defocusing force and will be deflected to large angles.

These particles can enter the acceptance of the innermost subdetectors and be the source of background.
To evaluate the level of this background, the IPC particles have been tracked in the CLD detector model~\cite{cld} using Key4HEP.
In particular, the occupancy in the vertex detector (VXD) and inner/outer trackers (TRK) has been studied.
The model description of the VXD barrel has been modified after the CDR since the study described in Ref.~\cite{cld}, following the reduction of the smaller central vacuum chamber from 15\,mm to 10\,mm.
The occupancy is defined as:

\begin{equation}
    occupancy = HD_{BX} \times S_{size}\times C_{size} \times S_{factor}
    \label{eq:occ}
\end{equation}

\noindent where $HD_{BX}$ is the hit density per bunch crossing (expressed in $hits/mm^2/BX$), $S_{size}$ is the sensor size ($25\,\mu$m$\times25\,\mu$m for the VXD silicon pixels, and $1\,$mm$\times0.05$\,mm for the TRK strips), $C_{size}$ is the cluster size (5 for the pixels and 2.5 for the strips), and $S_{factor}$ is a safety factor to account for uncertainties on the simulation of detector materials effects (e.g. backscattering), which for this study has been set to 3.

In a silicon detector, the occupancy is the number of channels that have signal above threshold.
To account for this, a cut on the energy deposited in the detectors has been set to 4\,keV for the VXD and 8\,keV for the TRK.

Table~\ref{tab:ipc_occ} summarizes the results of the study conducted for the four FCC-ee working points.
The number of $e^+e^-$ pairs produced increases proportionally with the beam energy, as expected from the cross section.
The maximum occupancy recorded for each subdetector is also larger at higher energies, but in particular for the vertex detector barrel it grows more rapidly with respect to the total number of pairs.
This can be explained with the fact that a larger fraction of the particles is produced inside the acceptance of the VXD at higher energies, as highlighted by the red area in Figure~\ref{fig:ipc_kin}.

\begin{table*}[h!bt]
\centering
\begin{tabular}{ r  |  cccc}
\hline \hline
   & Z & WW & ZH & Top \\
\hline 
Pairs produced per Bunch Crossing & $1300$ & $1800$ & $2700$ & $3300$ \\
Max occupancy VXD Barrel &  $70\times10^{-6}$ & $280\times10^{-6}$ & $410\times10^{-6}$ & $1150\times10^{-6}$ \\
Max occupancy VXD Endcap &  $23\times10^{-6}$ &  $95\times10^{-6}$ & $140\times10^{-6}$ &  $220\times10^{-6}$ \\
Max occupancy TRK Barrel &   $9\times10^{-6}$ &  $20\times10^{-6}$ &  $38\times10^{-6}$ &   $40\times10^{-6}$ \\
Max occupancy TRK Endcap & $110\times10^{-6}$ & $150\times10^{-6}$ & $230\times10^{-6}$ &  $290\times10^{-6}$ \\
\hline
Bunch Spacing [ns] & 30 & 345 & 1225 & 7598 \\
Max occ. VXD w/ $1\,\mu$s pileup  & $2.33\times10^{-3}$ & $0.81\times10^{-3}$ & --- & --- \\
Max occ. VXD w/ $10\,\mu$s pileup & $23.3\times10^{-3}$ & $8.12\times10^{-3}$ & $3.34\times10^{-3}$ & $1.51\times10^{-3}$ \\
 \hline \hline
\end{tabular}
 \caption{Pairs produced per bunch crossing at the four FCC-ee working points and maximum occupancy in the VXD and TRK subdetectors of CLD, also considering the pileup effect in two arbitrary readout time windows of $1\mu$s and $10\mu$s.} \label{tab:ipc_occ}
\end{table*}

The detector group poses a constraint to the allowable occupancy, set to a limit of about $1\%$.
Table~\ref{tab:ipc_occ} shows that in no subdetector the occupancy exceeds this safety value, even when accounting for the possibility of pileup in a hypothetical electronics readout window of $1\,\mu$s.

\begin{figure*}[!hbt]
    \centering
    \begin{subfigure}{0.49\textwidth}
    	\centering
    	\includegraphics[width=0.9\textwidth]{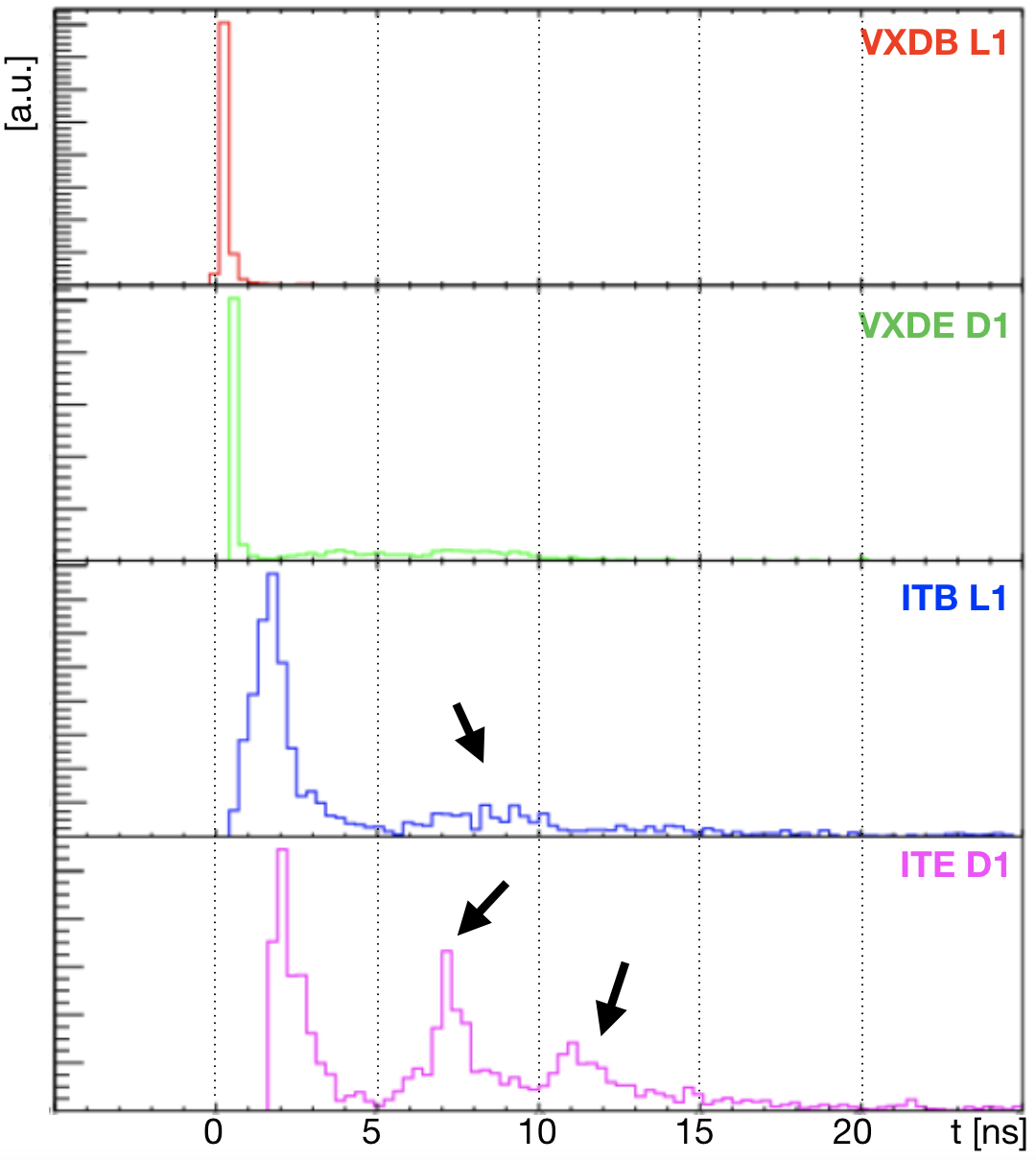}
    \end{subfigure}
    \begin{subfigure}{0.49\textwidth}
    	\centering
    	\includegraphics[width=0.9\textwidth]{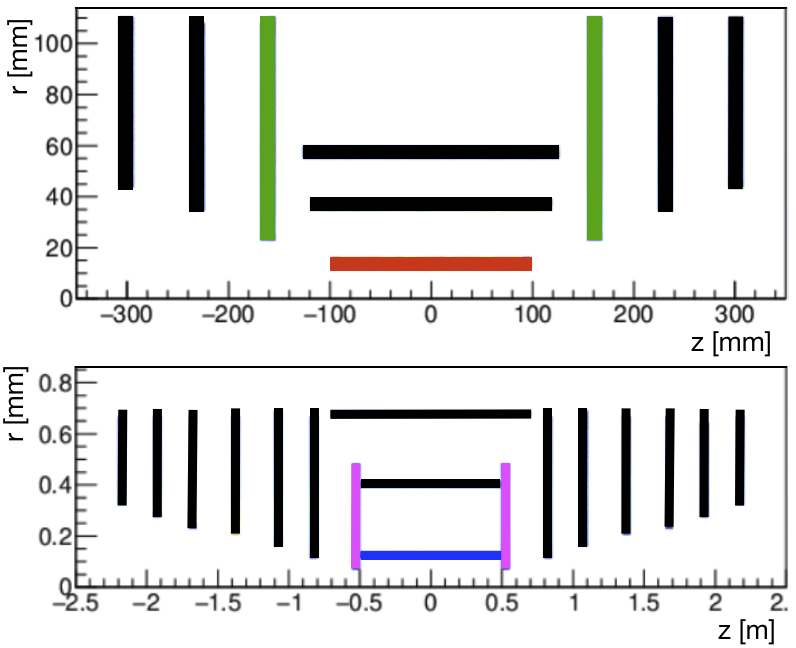}
    \end{subfigure}
    \caption{Left: arrival time of the particles in the subdetectors. Right: Position of the considered subdetectors: VXDB L1 (red), VXDE D1 (green), ITB L1 (blue), ITE D1 (magenta).}
    \label{fig:ipc_time}
\end{figure*}

The arrival time of the particles at the innermost subdetectors of the CLD vertex detector and inner tracker (IT) is shown on the left side of Figure~\ref{fig:ipc_time}.
The position of these elements is shown in the right side of Figure~\ref{fig:ipc_time}, respectively VXD Barrel Layer 1 (VXDB L1 - red), VXD Endcap Disk 1 (VXDE D1 - green), IT Barrel Layer 1 (ITB L1 - blue), IT Endcap Disk 1 (ITE D1 - magenta).
The first peak in the arrival time distribution corresponds to particles travelling straight from the IP to the detector, but - in particular for the inner tracker - secondary peaks can be observed after the first one.
These can be explained as a contribution from backscattered particles.
This signal could be rejected offline, further reducing the background induced by the IPC.

A consideration that should be done by looking at Equation~\ref{eq:occ} is that it strongly depends on the sensor and cluster size.
While for a Silicon based detector these values - in particular the sensor size - can be very small, they can be much larger for other types of detectors (e.g. drift chambers, wire chambers, ...) and the same background source can be negligible for a detector but relevant for another.
Therefore it is important to repeat these studies for the other experiment proposals for FCC-ee.

\section{\label{sec:conclusions} Conclusion}

We have discussed the properties of the beamstrahlung radiation emitted in a single beam-beam interaction for the FCC-ee, showing that we estimate an intense power of about 500 kW for v530 optics.
We showed that the beamstrahlung photons hit the vacuum chamber inside the first dipole magnet downstream the IP, at about 60\,m from the IP.
This results is the basis for the design of a dedicated extraction line to transport the beamstrahlung radiation up to a photon beam dump.
Based on these first evaluations, detailed FLUKA simulations and dedicated study to design the magnets and beam pipe for the extraction line are planned, to develop a concept for the photon beam dump including the proper target material and finally to design the alcove for the dump. 
This work is part of the FCC-ee MDI study group.

We have described the induced backgrounds in the CLD detector due to beamstrahlung radiation.
Detector backgrounds simulations show that the occupancy in the CLD Silicon vertex and tracker detectors due to the incoherent pairs produced by the beamstrahlung photons at the IP is below the safety value set by the detector group of $1\%$, also accounting for hypothetical pileup effects.
The analysis of the background particles arrival time showed a non-negligible contribution coming from backscattering, which could be removed offline further reducing the noise.
We plan to repeat the study for the other FCC-ee detectors, as different detector technologies can significantly affect the final occupancy.

\section{Acknowledgments}
The authors would like to thank the colleagues from the MDI group for useful discussions, in particular 
Helmut Burkhardt, 
Anton Lechner, Emmanuel Francois Perez, 
Dmitry Shatilov, 
and Katsunobu Oide. 

\section{Funding}
This work was partially supported by the European Union’s Horizon 2020 research and innovation programme under grant No 951754 – FCCIS Project.

\section{Declarations}
The authors declare that they have no competing interests.


\begin{thebibliography}{99}

\bibitem{fcc}
A.~Abada {\it et al.} [FCC Collaboration],
\newblock FCC-ee: The Lepton Collider : Future Circular Collider Conceptual Design Report Volume 2.
\newblock Eur.\ Phys.\ J.\ ST {\bf 228} (2019) no.2,  261.
\newblock   https://fcc.web.cern.ch/

\bibitem{guineapig}
D.~Schulte,
\newblock Beam-Beam Simulations with GUINEA-PIG .
CERN-PS-99-014-LP,
http://cds.cern.ch/record/382453, Mar, 1999.

\bibitem{key4hep}
G.~Ganis {\it et al.},
\newblock Key4hep, a framework for future HEP experiments and its use in FCC.
\newblock arXiv:2111.09874 [hep-ex]

\bibitem{yoyoka}
K.~Yokoya, P.~Chen, 
\newblock Beam-Beam Phenomena in Linear Colliders. 
Lect. Notes Phys. \textbf{400} (1992) 415-445

\bibitem{Sands:1969lzn}
M.~Sands,
\newblock The Physics of Electron Storage Rings: An Introduction.
Conf. Proc. C \textbf{6906161} (1969), 257-411,
SLAC-R-121.

\bibitem{prab_dmitry}
A.~Bogomyagkov, E. Levichev, and D. Shatilov
\newblock Beam-beam effects investigation and parameters optimization for a circular e+e- collider at very high energies.
Phys. Rev. ST Accel. Beams 17, 041004

\bibitem{Garcia:2016IPAC}
M.~A.~V.~Garc\'\i{}a and F.~Zimmermann,
\newblock Effect of Beamstrahlung on Bunch Length and Emittance in Future Circular e+e- Colliders.
Proceedings of IPAC2016, Busan, Korea WEPMW010

\bibitem{Shatilov}
D.~Shatilov {\it et al.},
\newblock LIFETRAC Code for the Weak-Strong Simulation of the Beam-Beam Effects in Tevatron.
21st IEEE Particle Accelerator Conference, Knoxville, TN, USA, 16 - 20 May 2005, pp.4138

\bibitem{oide-fccis}
K.~Oide,
\newblock Collider Optics.
FCCIS 2022 Workshop 
https://indico.cern.ch/event/1203316/ - 06/12/2022

\bibitem{slc}
T.~Barklow {\it et al.},
\newblock Experimental evidence for beam-beam disruption at the SLC.
SLAC-PUB-8043

\bibitem{ilc}
G.~Aarons {\it et al.},
\newblock International Linear Collider Reference Design Report.

\bibitem{clic}
M.~Aicheler {\it et al.},
\newblock The Compact Linear Collider (CLIC) – Project Implementation Plan.
\newblock CERN Yellow Reports: Monographs, Vol. 4/2018, CERN–2018–010–M,
https://doi.org/10.23731/CYRM-2018-004

\bibitem{nlc}
T.~Raubenheimer {\it et al.},
\newblock Parameters of the SLAC Next Linear Collider.
DOI: 10.1109/PAC.1995.504762

\bibitem{skekb}
K.~Akai {\it et al.},
\newblock SuperKEKB Collider.
arXiv:1809.01958v2 [physics.acc-ph] 10 Sep 2018

\bibitem{crabw}
M.~Zobov et al.,
\newblock Test of “Crab-Waist” Collisions at the DA$\Phi$NE $\Phi$-Factory.
\newblock Phys. Rev. Lett. 104, \textbf{174801}

\bibitem{Garcia:2019nci}
M.~A.~V.~Garc\'\i{}a and F.~Zimmermann,
\newblock Beam blow up due to Beamstrahlung in circular $e^+e^-$ colliders.
Eur. Phys. J. Plus \textbf{136} (2021) no.5, \textbf{501},
doi:10.1140/epjp/s13360-021-01485-x
[arXiv:1911.03420 [physics.acc-ph]].

\bibitem{hbu_bs}
H.~Burkhardt,
\newblock Radiation Generated at the IP.
FCC-ee Optics Design Meeting \#105 https://indico.cern.ch/event/854159/

\bibitem{Boscolo:2021hsq}
M.~Boscolo, N.~Bacchetta, M.~Benedikt, L.~Brunetti, H.~Burkhardt, A.~Ciarma, M.~Dam, F.~Fransesini, M.~Jones and R.~Kersevan, \textit{et al.},
\newblock Challenges for the Interaction Region Design of the Future Circular Collider FCC-ee.
doi:10.18429/JACoW-IPAC2021-WEPAB029, [arXiv:2105.09698 [physics.acc-ph]].

\bibitem{jeremie}
C.J.~Eriksson, J. Bauche,
\newblock Magnet Design for Beamstrahlung Photons Extraction Line.
FCC-ee MDI meeting \#43 and FCCIS WP2.3 meeting \#14 
https://indico.cern.ch/event/1241377/

\bibitem{madx}
\newblock MAD-X, Methodical Accelerator Design.
\newblock http://mad.web.cern.ch/mad

\bibitem{fani}
F.~Valchkova-Georgieva,
\newblock Tunnel Integration.
FCCIS 2022 Workshop, 
https://indico.cern.ch/event/1203316/ - 06/12/2022

\bibitem{v566}
K.~Oide,
\newblock Layout, optics, parameters.
FCCWeek 2023, London, 
https://indico.cern.ch/event/1202105/

\bibitem{BBBREM}
R.~Kleiss, H.~Burkhardt,
\newblock BBBREM – Monte Carlo simulation of radiative Bhabha scattering in the very forward direction.
\newblock CERN SL/94-03 (OP) January 1994

\bibitem{rb_dafne}
M.~Greco, G.~Montagna, O.~Nicrosini, F.~Piccinini,
\newblock Radiative Bhabha scattering at DAΦNE.
\newblock Physics Letters B, Vol. \textbf{318}, Issue 4, 16 December 1993, Pages 635-641

\bibitem{rb_vepp}
A.F.~Blinov \textit{et al.},
\newblock Luminosity measurement with the MD-1 detector at VEPP-4.
\newblock Nucl. Instr. and Meth. in Physics Research \textbf{A273} (1988) 31-39

\bibitem{rb_pep2}
S.~Ecklund, C.~Field, G.~Mazaheri,
\newblock A Fast Luminosity Monitor System for PEP II.
\newblock SLAC-PUB-8688 October 30, 2000

\bibitem{cld}
N.~Bacchetta \it{et al.},
\newblock CLD - A Detector Concept for FCC-ee.
arXiv:1911.12230 [physics.ins-det]

\end{thebibliography}
\end{document}